TEL AVIV UNIVERSITY

# Frame Codes for the Block-Erasure Channel

by
Itamar Jacoby

Supervisor:
Prof. Ram Zamir

## ABSTRACT

Analog codes add redundancy by expanding the dimension using real/complex-valued operations. Frame theory provides a mathematical basis for constructing such codes, with diverse applications in non-orthogonal code-division multiple access (NOMA-CDMA), distributed computation, multiple description source coding, space-time coding (STC), and more. The channel model corresponding to these applications is a combination of noise and erasures. Recent analyses showed a useful connection between spectral random-matrix theory and large equiangular tight frames (ETFs) under random uniform erasures. In this work we generalize this model to a channel where the erasures come in blocks. This particularly fits NOMA-CDMA with multiple transmit antennas for each user and STC with known spatial grouping. We present a method to adjust ETF codes to suit block erasures, and find minimum intra-block-correlation frames which outperform ETFs in this setting.

# Contents

# I. Introduction

A frame $F$ is a $M \times N$ matrix $(N > M)$ that consists of $N$ vectors that form an over-complete basis over a space of dimension $M$ [29]. Frame codes utilize $N$ vectors of length $M$ taken from a frame $F$ for either adding "analog" redundancy to source information or multiplexing information over a common resource, and are especially suitable for channels which combine noise and erasures [9], [8], [24], [16]. Applications of frame codes include non-orthogonal code-division multiple access (NOMA-CDMA) [33], [34], [26], multiple-input multiple-output (MIMO) beamforming [20], space-time coding (STC) [11], [15], multiple-description (MD) source coding [9], [7], [22], [6], coded distributed computation [32], and more.

Recent works investigated the asymptotic behavior of frames, especially of equiangular tight frames (ETFs), and showed that for large ETFs the singular values of a randomly selected subframe are distributed according to the MANOVA distribution [12], [13], [14], [11], [21], [19]. These as well as other works demonstrated applications of ETFs for different problems involving erasures, and provided empirical evidence for ETF superiority over other frame families [11], [6], [23], [10], [32]. These works assumed a combinatorically symmetric "N choose K" erasure channel, which is equivalent to random uniform selection of a subframe of size $M \times K$. The well- or ill-conditioning of this subframe determines the system performance.

In this work we present a combinatorial *block-erasure* channel, which generalizes the uniform erasure model. In the block-erasure channel, erasures are performed over entire blocks of vectors jointly, introducing constraints on the selection of the subframes. In this model the size and position of the blocks are predefined, unlike in the classic burst-erasure channel [5], [17]. Such a setting is especially interesting in the context of NOMA-CDMA with multiple-antenna users, or STC with known spatial grouping of the transmit antennas, and to the best of our knowledge has not been investigated previously in the context of frame codes. While ETFs are invariant to frame permutation for a uniform selection probability of the subframes, for block erasures this is not the case. This raises the following questions: what is the performance of ETF for this problem setting? Can it be improved by properly allocating vectors to each block? And can we find new frames that are better than ETF for the block-erasure channel?

We begin with a brief overview of frame theory and some ETF properties in Section II. In Section III we define the block-erasure channel, demonstrate it in the NOMA-CDMA and STC setups, and present theoretical bounds for these setups. In Section IV we present two conjectures supported by empirical evidence: one regarding the structure of the squared correlation of optimal block-erasure frames, and the other regarding a required arrangement of ETF columns in order to maintain the superior MANOVA spectrum. In Section V we present the performance of the suggested block-erasure frames for both the NOMA-CDMA and STC setups. The newly designed frame codes achieve higher average capacity in the NOMA-CDMA setup, and lower error probability in the STC setup, when compared with ETFs and near-ETFs. They also significantly reduce the capacity outage probability, thus providing a more reliable communication. These results were presented in 2024 IEEE International Symposium on Information Theory (ISIT) [18].

# II. Background

## A. Frame Theory

A matrix $\boldsymbol{F}$ of size $M \times N$, composed of a set of $N$ vectors $\{\boldsymbol{f_n}\}_{n=1}^{N}$ in a Hilbert space $H$ of dimension $M, N > M,$ is considered a frame if there exist constants $a, b > 0$ such that

$$a||\boldsymbol{x}||^2 \leq \sum_{n=1}^{N} |<\boldsymbol{x}, \boldsymbol{f_n}>|^2 \leq b||\boldsymbol{x}||^2 \tag{1}$$

for every vector $\boldsymbol{x} \in H$. In case $a = b$, $F$ is considered a tight frame. Furthermore, if $a = b = N/M$, then $F$ is considered a unit tight frame (UNTF) and is composed of unit-normalized vectors. The ratio $N/M$ is called the frame redundancy. The correlation between vectors of a frame is defined as

$$c_{n,k} = <\boldsymbol{f_n}, \boldsymbol{f_k}> = \boldsymbol{f_n^H f_k}, \qquad 1 \leq n, k \leq N. \tag{2}$$

The Welch bound defines a lower bound on the averaged squared correlation between all distinct pairs of vectors [30],

$$\frac{1}{(N-1)N}\sum_{n\neq k}^{N}\left|c_{n,k}\right|^2 \geq \frac{N-M}{(N-1)M}, \tag{3}$$

and is achieved with equality if and only if $F$ is UNTF. This also implies the following maximum Welch bound,

$$\max_{1\leq n\neq k\leq N}\left|c_{n,k}\right|^2 \geq \frac{N-M}{(N-1)M} \triangleq \epsilon_{WB}, \tag{4}$$

Which is achieved with equality if and only if $F$ is ETF. This yields the following property of ETFs:

$$\left|c_{n,k}^{ETF}\right|^2 = \begin{cases} 1, & n = k \\ \epsilon_{WB}, & n \neq k \end{cases}, \tag{5}$$

## *B. ETF Generation by Difference Sets*

A common method to generate ETFs is related to *difference sets*. A subset $D$ of size $M$ of a finite abelian group $G$ of size $N$ is a $(N, M, \lambda)$ difference set in $G$ if all non-zero elements in $G$ can be expressed as the difference of some two members of $D$ exactly $\lambda$ times. This is possible only for specific combinations of $(N, M, \lambda)$ that satisfy $\lambda = M(M-1)/(N-1)$ (a necessary but not sufficient condition) [2].

*Harmonic frames* are constructed by selecting any set of $M$ rows from a $N \times N$ DFT matrix. In case the set of rows form a difference set over $(\mathbb{Z}_N, +)$ the frame is a harmonic ETF [31]. Similarly, *Hadamard ETFs* are constructed by selecting a set of $M$ rows from a $N \times N$ unitary Hadamard matrix $(N = 2^L)$ which form a difference set over $GF(2^L, +)$ [3], [4].

## *C. Gram Matrix and Eigenvalue Asymptotic Distribution*

The Gram matrix of a frame $F$ is defined as

$$G = F^H F, \tag{6}$$

with elements equal to $c_{n,k}$ (2). Let $F$ be an ETF, $S_K$ be a K-subset of $\{1, \ldots, N\}$, and $F_{S_K}$ be a subframe of $F$ of size $M \times K$ constructed by the columns defined by $S_K$. The Gram matrix of $F_{S_K}$ is defined as

$$G_{S_K} = F_{S_K}^H F_{S_K}. \tag{7}$$

It was shown empirically in [13], and proved mathematically in [21], [11], [14] and [19], that for a random uniform selection of $S_K$, the distribution of the eigenvalues of $G_{S_K}$ converges as $N \to \infty$ to the MANOVA distribution with parameters $\beta$ and $\gamma$:

$$\lim_{M,N\to\infty} f_\lambda(\lambda) = f_{\beta,\gamma}^{MANOVA} = \frac{\sqrt{(\lambda_+ - \lambda)(\lambda - \lambda_-)}}{2\pi\beta\lambda(1-\gamma\lambda)} \mathbb{1}_{[\lambda_-,\lambda_+]}(\lambda) + \left(1 + \frac{1}{\beta} - \frac{1}{\beta\gamma}\right)^+ \delta\left(\lambda - \frac{1}{\gamma}\right) \tag{8}$$

where $\lambda_\pm = \left(\sqrt{\beta(1-\gamma)} \pm \sqrt{1-\beta\gamma}\right)^2$, $\beta = \lim K/M$ and $\gamma = \lim M/N$.

Similar asymptotic behavior holds also for various families of non-ETF frames, including harmonic or Hadamard frames with irregular spectrum [13], [6], [10]. These “near-ETF” frames satisfy relaxed requirements on the squared correlation or the difference set spectrum compared to (5). This allows constructing frames with a wider pool of $(M, N)$ combinations.

# III. Block-Erasure Channel Definition and Use Cases

Our work deals with communication systems that utilize frame codes for data transmission over noisy block-erasure channels. We begin by defining the block-erasure channel.

## A. Block-Erasure Channel

In the block-erasure problem, a general frame $F$ is divided into $N_B$ blocks, each with $N_v = N/N_B$ vectors. We assume a consecutive division of the frame,

$$F = [\boldsymbol{f_1}, \boldsymbol{f_2}, \dots, \boldsymbol{f_N}] = \left[F_{B_1}, F_{B_2}, \dots, F_{B_{N_B}}\right], \tag{9}$$

where $F_{B_i}$ are the blocks submatrices

$$F_{B_i} = [\boldsymbol{f}_{(i-1)N_v+1}, \boldsymbol{f}_{(i-1)N_v+2}, \dots, \boldsymbol{f}_{(i-1)N_v+N_v}]. \tag{10}$$

In the presence of erasures, $N_E$ random blocks are erased to a total of $N - K = N_E N_v$ erased column vectors. We are left with a subframe $F_{S_K}$ of size $M \times K$ composed of $N_A = N_B - N_E > 1$ "active" (non-erased) blocks, and a total of $K = N_A N_v$ active vectors. This reduces to the uniform erasure case of [23], [10], [11] if $N_B = N$ and $N_A = K$, meaning that each column vector is in a separate block ($N_v = 1$).

We define the following ratios that are relevant for asymptotic analysis: the non-erasure probability $p$, the aspect ratio of the frame $\gamma$ and the aspect ratio of the subframe $\beta$ [11],

$$p = \frac{K}{N} = \frac{N_A}{N_B}, \qquad \gamma = \frac{M}{N}, \qquad \beta = \frac{K}{M}. \tag{11}$$

We next discuss how this setting applies to the following setups: multi-user network employing NOMA-CDMA, and single user employing STC under erasure channel [1].

## B. The NOMA-CDMA Setup

In a NOMA-CDMA system we have a total of $N$ transmit antennas that belong to $N_B$ users, each equipped with $N_v$ antennas. Each user is provided with a block of $N_v$ spreading sequences, $F_{B_i}$ (10). Out of $N_B$ users, only $N_A$ are active and are transmitting simultaneously in an uncoordinated manner, meaning users have no knowledge of the number and identity of the other transmitting users. In total, $K = N_A N_v$ antennas are active. The system is designed such that $K \leq M < N$. Assuming additive white Gaussian channel (AWGN), the received signal is

$$\boldsymbol{y} = \sum_{n \in S_K} \boldsymbol{f_n} + \boldsymbol{z}, \tag{12}$$

where $S_K$ is the set of active antennas of all users and $\boldsymbol{z} \sim \mathcal{CN}(\boldsymbol{0}, \sigma_z^2 I_M)$ is the noise vector.

Because the set of codewords transmitted simultaneously in each time occasion are random, the total capacity can be described as a random variable $C$ having some distribution. Therefore, we measure the performance of such a system using the average capacity [28], [27]:

$$\mathbb{E}_{S_K}[C] \;=\; \mathbb{E}_{S_K}\left[\log_2\left(\det\left(I + SNR \cdot {F_{S_K}}^H F_{S_K}\right)\right)\right] = \mathbb{E}_{S_K}\left[\log_2 \prod_{k=1}^{K} (1 + SNR \cdot \lambda_k)\right] \tag{13}$$

where $\mathbb{E}_{S_K}[\cdot]$ denotes expectation over the selection of the subset $S_K \subset \{1, \dots, N\}$, $F_{S_K}$ is the subframe of size $M \times K$ constructed from the $K$ active codewords $\{f_n\}_{n \in S_K}$, $\{\lambda_k\}_{k=1}^{K}$ are the eigenvalues of $G_{S_K}$ (7) and $SNR = 1/\sigma_z^2$ is the signal to noise ratio. Because the unit-norm assumption implies $\frac{1}{K}\sum_{k=1}^{K} \lambda_k = 1$, the arithmetic-geometric means inequality implies

$$\left[\prod_{k=1}^{K} (1 + SNR \cdot \lambda_k)\right]^{1/K} \leq 1 + SNR, \tag{14}$$

And therefore (13) is upper bounded by

$$\mathbb{E}_{S_K}[C] \leq K \log_2(1 + SNR), \tag{15}$$

with equality if and only if the eigenvalues of $G_{S_K}$ are identical for all choices of $S_K$ [11]. This amounts to orthogonality for all choices of a subframe, which cannot be achieved because a frame is an *over-complete* basis. We call $K\log_2(1 + SNR)$ the *capacity orthogonality bound* which is unachievable for uniform random erasures. We will use the block-erasure constraints in our advantage to design frames with low eigenvalue spread, in the sense of low log arithmetic-to-geometric means ratio, which is equivalent to small KL-distance between uniform and the actual eigenvalue distribution, hence high average capacity. This is demonstrated by examining the gap between the average capacity (13) and the orthogonality bound (15),

$$C_{gap} = K\mathbb{E}_{S_K}\left[\log_2 \frac{(1+SNR)}{\left(\prod_{k=1}^{K}(1+SNR\cdot\lambda_k)\right)^{\frac{1}{K}}}\right]. \tag{16}$$

We define $\lambda_{S_K}(k) = \lambda_k$ as the eigenvalues of a given subframe $F_{S_K}$ of size $M \times K$ and $\lambda_U(k) = 1\ \forall k$. Then, by using arithmetic-geometric means inequality, and considering $\lambda_U(k) = 1\ \forall k$,

$$\begin{aligned} &\log_2 \frac{(1+SNR)}{\left(\prod_{k=1}^{K}(1+SNR\cdot\lambda_k)\right)^{\frac{1}{K}}} \\ &= \log_2 \frac{\frac{1}{K}\sum_{k=1}^{K}\left(1+SNR\lambda_U(k)\right)}{\left(\prod_{k=1}^{K}\left(1+SNR\lambda_{S_K}(k)\right)\right)^{\frac{1}{K}}} \underset{\substack{AM-GM \\ \lambda_{U(k)}=1\,\forall K}}{=} \log_2 \frac{\left(\prod_{k=1}^{K}\left(1+SNR\lambda_U(k)\right)\right)^{\frac{1}{K}}}{\left(\prod_{k=1}^{K}\left(1+SNR\lambda_{S_K}(k)\right)\right)^{\frac{1}{K}}} \\ &= \frac{1}{K}\log_2 \prod_{k=1}^{K}\frac{\left(1+SNR\lambda_U(k)\right)}{\left(1+SNR\lambda_{S_K}(k)\right)} = \sum_{k=1}^{K}\frac{1}{K}\log_2\left(\frac{1+SNR\lambda_U(k)}{1+SNR\lambda_{S_K}(k)}\right) \end{aligned} \tag{17}$$

We define the following as the eigenvalue distributions scaled by SNR:

$$\lambda_{S_K}^{SNR}(k) = \frac{1+SNR\lambda_{S_K}(k)}{K(1+SNR)}, \qquad k = 1, \dots, K, \tag{18}$$

$$\lambda_U^{SNR}(k) = \frac{1+SNR\lambda_U(k)}{K(1+SNR)} = \frac{1}{K}, \qquad k = 1, \dots, K. \tag{19}$$

Then by plugging (18) and (19) in (17), we can express the gap in (16) as the conditional relative entropy between these distributions.

$$\frac{1}{K}C_{gap} = \mathbb{E}_{S_K}\left[\sum_{k=1}^{K}\lambda_U^{SNR}(k)\log_2\left(\frac{\lambda_U^{SNR}(k)}{\lambda_{S_K}^{SNR}(k)}\right)\right] = D\left(\lambda_U^{SNR}, \lambda_{S_K}^{SNR} | S_k\right) \tag{20}$$

For asymptotic frames, the eigenvalues of any selected subframe are distributed according to the asymptotic frame distribution $\lambda_F^{SNR}$ (assuming such limit exist), and we can omit the expectancy and receive,

**Proposition 1:**

$$\frac{1}{K}C_{gap} \underset{N,M,K\to\infty}{=} D_{KL}\left(\lambda_U^{SNR}, \lambda_{S_K}^{SNR}\right). \tag{21}$$

## *C. The STC Setup*

In the context of STC, we have a single user that is transmitting a space-time code $X$ from $N$ antennas over $M$ time slots. There are $L$ possible codewords, $C_1, C_2, \dots C_L$. Each codeword is a $M \times N$ matrix. We examine the case of $L = 2$, where there are two codewords in the codebook:

$$C_1 = 0.5F, \qquad C_2 = -0.5F, \tag{22}$$

and the difference between $C_1$ and $C_2$ is the frame $F$. We follow a similar division of the antennas to $N_B$ blocks out of which $N_E$ antenna blocks are shadowed or experience failures, and their transmission is erased for all $M$ time slots [1]. Each block of antennas consists of $N_v$ antennas, so we have $N - K$ erased antennas overall. The transmitted signal from the remaining $K$ antennas undergoes a Rayleigh fading channel with AWGN, and the channel is assumed to be constant for a period of $M$ time slots and independent between antennas. There is no knowledge at the transmitter of the channel or the erasures. The effective distance of the transmitted codeword can be described by the subframe $F_{S_K}$ of size $M \times K$ $(M \leq K < N)$, whose $M \times M$ Gram matrix is given by (7). Since the erasures are random, we use the average error probability as the performance measure. Assuming maximum likelihood (ML) decoder, the average error probability $p_e$ of such a system is upper bounded by [27], [25]:

$$p_e \le p_e^{Bound} = \mathbb{E}_{S_K}\left[\frac{1}{\prod_{k=1}^{M}\left(1+\frac{SNR}{4}\cdot\lambda_k\right)}\right] \quad (23)$$

where $\lambda_k$ are eigenvalues of $G_{S_K}$ as in (13). We can apply the arithmetic-geometric means inequality in order to find a lower bound on the the average error probability bound $p_e^{Bound}$:

**Proposition 2:**

$$p_e^{Bound} \ge \mathbb{E}_{S_K}\left[\frac{1}{\left(\frac{1}{M}\sum_{k=1}^{M}\left[1+\frac{SNR}{4}\cdot\lambda_k\right]\right)^M}\right] = \frac{1}{\left(1+\frac{SNR}{4}\right)^M}. \quad (24)$$

We name the last term as the *error probability orthogonality bound*. Similarly to the bound on (15), it is achieved with equality if and only if all eigenvalues are identical for all choices of $S_K$, which is not feasible as was previously mentioned. Hence, like for the average capacity (15), a criterion for low average error probability for space-time frame codes for the block-erasure channel is to have low eigenvalue spread for all achievable subframes under the block-erasure constraints, which is equivalent to small KL-distance between uniform and the actual eigenvalue distribution (21).

# IV. Properties of Block-Erasure Frames

In this section we propose two conjectures. The first is regarding the desired correlation structure of a frame under block erasures, which differs from the structure of ETF (that was shown empirically to be superior in the non-block case). The second is regarding the eigenvalue distribution the Gram matrix of subframes under block erasures. We support both conjectures by empirical results obtained from frames optimized for the block erasure channel.

## A. Frame Construction

In our analysis we limit ourselves to harmonic (DFT-based) frames [31] or Hadamard-based frames [4]. We construct the frames by selecting $M$ rows of the $N \times N$ DFT or Hadamard matrix, and by permuting the columns. We use the following notation:

$$F = Z_S W P, \quad (25)$$

where $W$ is a DFT matrix or a normalized Hadamard matrix, $S = \{s_1, s_2, \dots, s_M\}$ is the set of $M$ selected rows, $Z_S$ is a $M \times N$ row-selection matrix that has a single non-zero value equal to 1 in each row, in a column defined by the set $S^M$, and $P$ is a permutation matrix. We use a permutation vector $\boldsymbol{\pi}$ of length $M$ to mark the non-zero indices in each column of $P$.

We search for optimal frames for the block erasures that would maximize the average capacity for the NOMA-CDMA setup (13) using exhaustive search. First, for a given difference set of ETF or almost-difference set based near-ETF frames [2], we look for optimal column permutation. Optimal frames of this design are referred as *Permuted ETF (PETF)*. We also look for new frames with optimal design for this problem, constructed by an optimal set and an optimal permutation. These frames are referred to as *Block Unit Tight Frames (BUTF)*. Using the notation from (25),

$$F_{PETF} = Z_{S_{ETF}} W P^*, \quad (26)$$

$$F_{BUTF} = Z_{S^*} W P^*, \quad (27)$$

where $S_{ETF}$ is a difference set or an almost-difference set, and $S^*$ and $P^*$ are the maximizers of the average capacity (13). We have shown in Section III that both the average capacity bound for NOMA-CDMA and the average error probability bound for STC get closer to their respective orthogonality bounds by having a low eigenvalue spread.

## B. Desired Correlation Structure

Let us examine the $N \times N$ squared correlation matrix of a general frame, $(F^H F)^{\circ 2} = G \circ G$, where $(\circ)$ represents the element-wise product. The elements of this matrix are $\left|c_{n,k}\right|^2$ (2). We examine block matrices of size $N_v \times N_v$ within of this matrix denoted by $B_{g,l},\ g, l = 1, \dots, N_B$. Block matrices around the diagonal $(l = g)$ contain the intra-block correlation of vectors $\boldsymbol{f_n}$ in each block, and the other block matrices $(l \neq g)$ contain the inter-block correlation of the vectors. We provide in (28) an example of this notation with 4 blocks.

$$(F^H F)^{\circ 2} = \begin{bmatrix} \boldsymbol{B_{1,1}} & B_{1,2} & B_{1,3} & B_{1,4} \\ B_{1,2}^T & \boldsymbol{B_{2,2}} & B_{2,3} & B_{2,4} \\ B_{1,3}^T & B_{2,3}^T & \boldsymbol{B_{3,3}} & B_{3,4} \\ B_{1,4}^T & B_{2,4}^T & B_{3,4}^T & \boldsymbol{B_{4,4}} \end{bmatrix} \tag{28}$$

Under block erasures, only specific subframes are possible. For the example in (28) with $N_A = 2$ active blocks selected in random, there are $\binom{4}{2} = 6$ possible subframes with the following squared correlation matrices,

$$\left\{ \begin{matrix} \begin{bmatrix} \boldsymbol{B_{1,1}} & B_{1,2} \\ B_{1,2}^T & \boldsymbol{B_{2,2}} \end{bmatrix}, \begin{bmatrix} \boldsymbol{B_{1,1}} & B_{1,3} \\ B_{1,3}^T & \boldsymbol{B_{3,3}} \end{bmatrix}, \begin{bmatrix} \boldsymbol{B_{1,1}} & B_{1,4} \\ B_{1,4}^T & \boldsymbol{B_{4,4}} \end{bmatrix} \\ \begin{bmatrix} \boldsymbol{B_{2,2}} & B_{2,3} \\ B_{2,3}^T & \boldsymbol{B_{3,3}} \end{bmatrix}, \begin{bmatrix} \boldsymbol{B_{2,2}} & B_{2,4} \\ B_{2,4}^T & \boldsymbol{B_{4,4}} \end{bmatrix}, \begin{bmatrix} \boldsymbol{B_{3,3}} & B_{3,4} \\ B_{3,4}^T & \boldsymbol{B_{4,4}} \end{bmatrix} \end{matrix} \right\}. \tag{29}$$

As can be seen from (29), the probability of appearance of the intra-block correlation matrices is higher compared to the inter-block correlation matrices. If so, can we trade lower correlation values of elements in the intra-block correlation matrices for higher correlation values than $\epsilon_{WB}$ (4) for some of the elements in the inter-block correlation matrices?

Indeed, we observe empirically that UNTFs optimized for block erasures have a squared correlation that differs from that of ETF (5), and approaches the following desired structure:

$$\left|c_{n,k}^{desired}\right|^2 = \begin{cases} 1, & n = k \\ 0, & n \neq k, c_{n,k} \in B_{g,g} \\ \epsilon, & c_{n,k} \notin B_{g,g}, \end{cases} \tag{30}$$

where the value of $\epsilon$ is computed to satisfy the average Welch bound (3), which is achieved with equality for UNTFs [30],

$$\epsilon = \left(1 + \frac{N_v - 1}{N_v(N_B - 1)}\right)\epsilon_{WB}. \tag{31}$$

While the structure in (30) is most likely unachievable for most UNTFs defined by the tuple $(M, N, N_B)$, the intra-block correlation matrices $B_{g,g}$ tend to an identity matrix and have lower correlation compared to the inter-block correlation matrices. Such a design also hints at a low eigenvalue spread, which is required for good performance as discussed in Section III. We support this claim with empirical observations demonstrated in Figure 1, which shows the distance between the intra-block correlation matrices of a given frame and the identity matrix, where the distance measure is defined by:

$$d = \sqrt{\sum_g \sum_{n,k} \left(B_{g,g}(n,k) - I_{N_V}(n,k)\right)^2}. \tag{32}$$

We see that Hadamard BUTFs (27), optimized for the block-erasure channel have intra-block correlation matrices more similar to the identity matrix than Hadamard near-ETF. We also provide an example below.

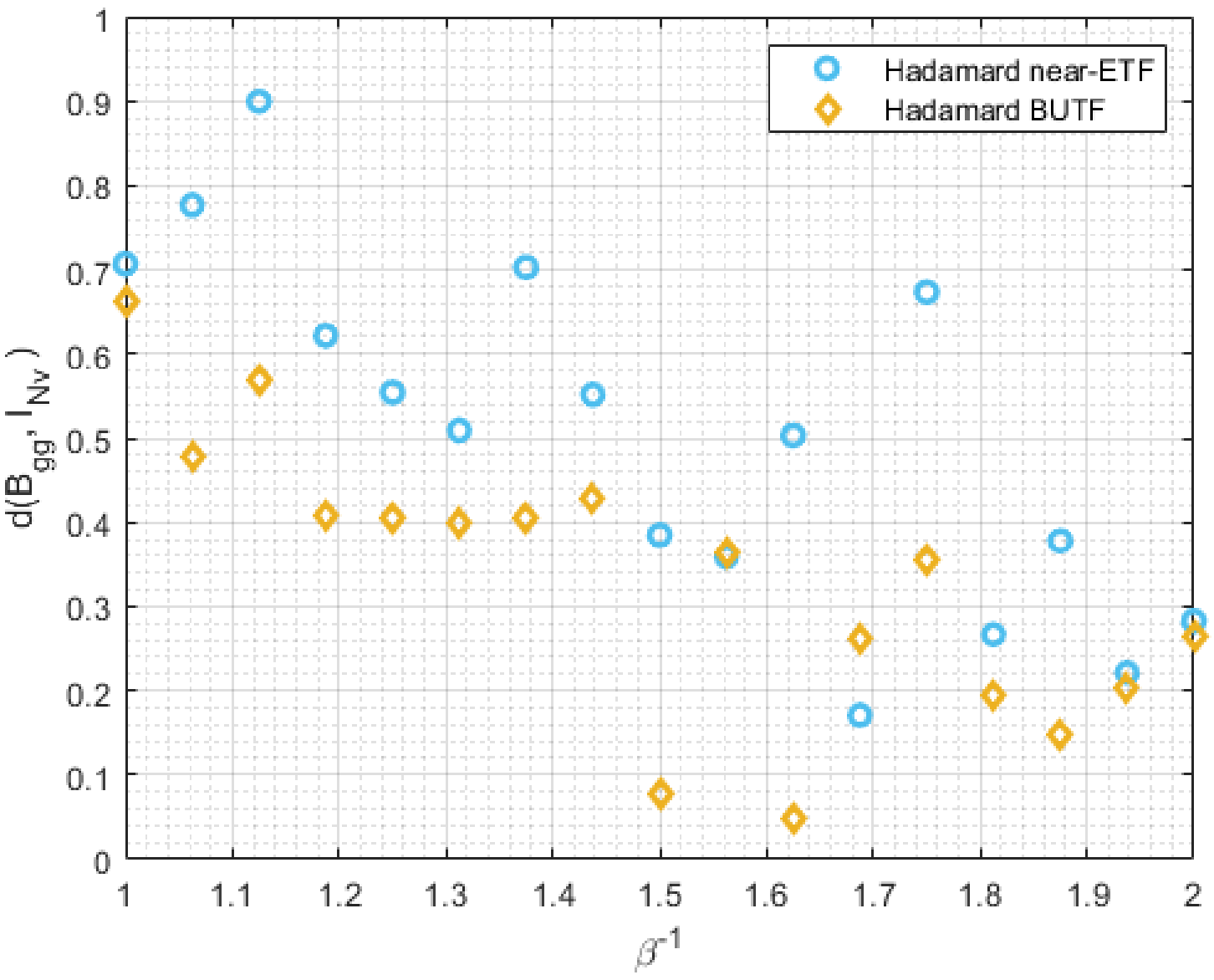

Figure 1: Distance of intra-block correlation matrices from identity matrix for $N = 64, N_B = 16, p = 0.25$

## C. Case Study: Comparison of ETF, PETF and BUTF

In the following example we provide a comparison between three Hadamard-based frames: ETF, PETF and BUTF, for configuration of $(M, N, N_B) = (6,16,4)$ with $N_A = 2$ and SNR = 30 dB (for the sake of this example and due to the low frame dimensions we violate the $M \geq K$ requirement). ETF is constructed by selecting the difference set of rows {0, 2, 5 ,6, 14, 15} of the Hadamard matrix using the original columns order. We denote this unpermuted ETF as "canonical". PETF is constructed by the same set, but uses the column permutation $\boldsymbol{\pi}^* =$ [2, 4, 7, 11, 10, 9, 14, 3, 15, 13, 1, 5, 12, 8, 6,16]. Both have the same squared correlation matrix (Figure 2), but the latter outperforms the former in a NOMA-CDMA setup: ETF has an average capacity of 52.9 bits/sec while Permuted ETF has 60.3 bits/sec. In comparison, BUTF that uses the set $S^* =$ {3, 4, 5, 6, 7, 9, 13} and the column permutation $\boldsymbol{\pi}^* =$ [7, 6, 12, 9, 2, 1, 8, 15, 5, 4, 3, 13, 16, 14, 10, 11], has an average capacity of 61 bits/sec, and outperforms the two previous frames. Examining the squared correlation matrix of the BUTF (Figure 3), we see that in the intra-block correlation matrices $B_{gg}$, some of the values $|c_{n,k}|^2$ are equal to zero, at the expense of having some values of $|c_{n,k}|^2 > \epsilon_{WB}$ in the inter-block correlation matrices.

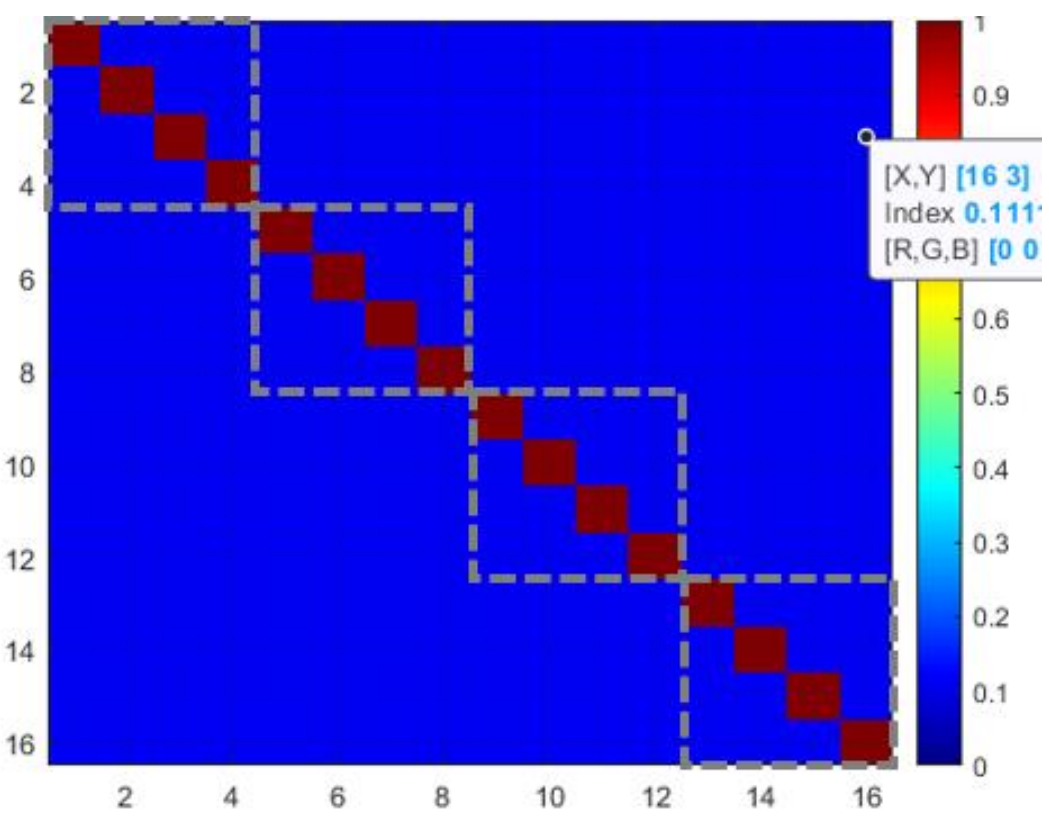

Figure 2: ETF/PETF squared correlation matrix for $N = 16, M = 6, N_B = 4$

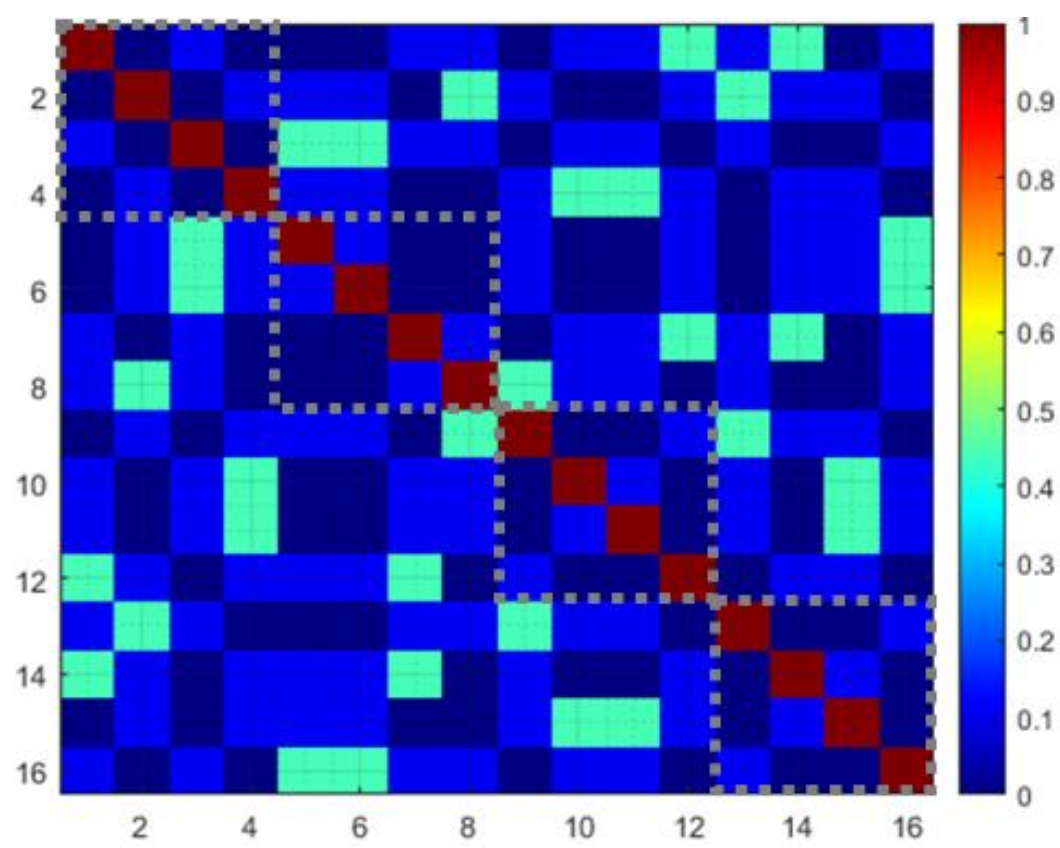

Figure 3: BUTF squared correlation matrix for $N = 16, M = 6, N_B = 4$

## D. Eigenvalue Distribution

As we saw in the previous example, for block erasures the order of the vectors that construct the frame is important. We show empirically that for canonical ETF and near-ETF frames the eigenvalue distribution of the Gram matrix of randomly selected subframes under block-erasure constraints shows a noncontinuous and sporadic behavior, which is very different from the superior MANOVA distribution achieved for uniform random erasures, [13], [21], [19]. We also show that proper allocation of the vectors into blocks by optimal column permutation provides a similar distribution to MANOVA. We support this claim using empirical analysis of the eigenvalue distribution of $G_{S_K}$ for Hadamard PETF and Hadamard BUTF, both constructed using optimal permutations. Results of this analysis is presented in Figure 4 and Figure 5.

In Figure 4 we compare PETF, BUTF and near-ETF with MANOVA distribution (8) and with Marchenko-Pastur (MP) distribution [28]. The latter fits the distribution of random i.i.d frames and serves as a reference. This figure shows an example of the distribution for a specific set of parameters, and we can see a close match for both PETF and BUTF with MANOVA. Figure 5 shows the Kullback–Leibler (KL) divergence between the eigenvalue distribution of the Gram matrices of the discussed frames and MANOVA distribution for different values of $\beta^{-1}$. We see that all PETF and most BUTF frames diverge very slightly from MANOVA. Canonical near-ETFs however diverge greatly from MANOVA.

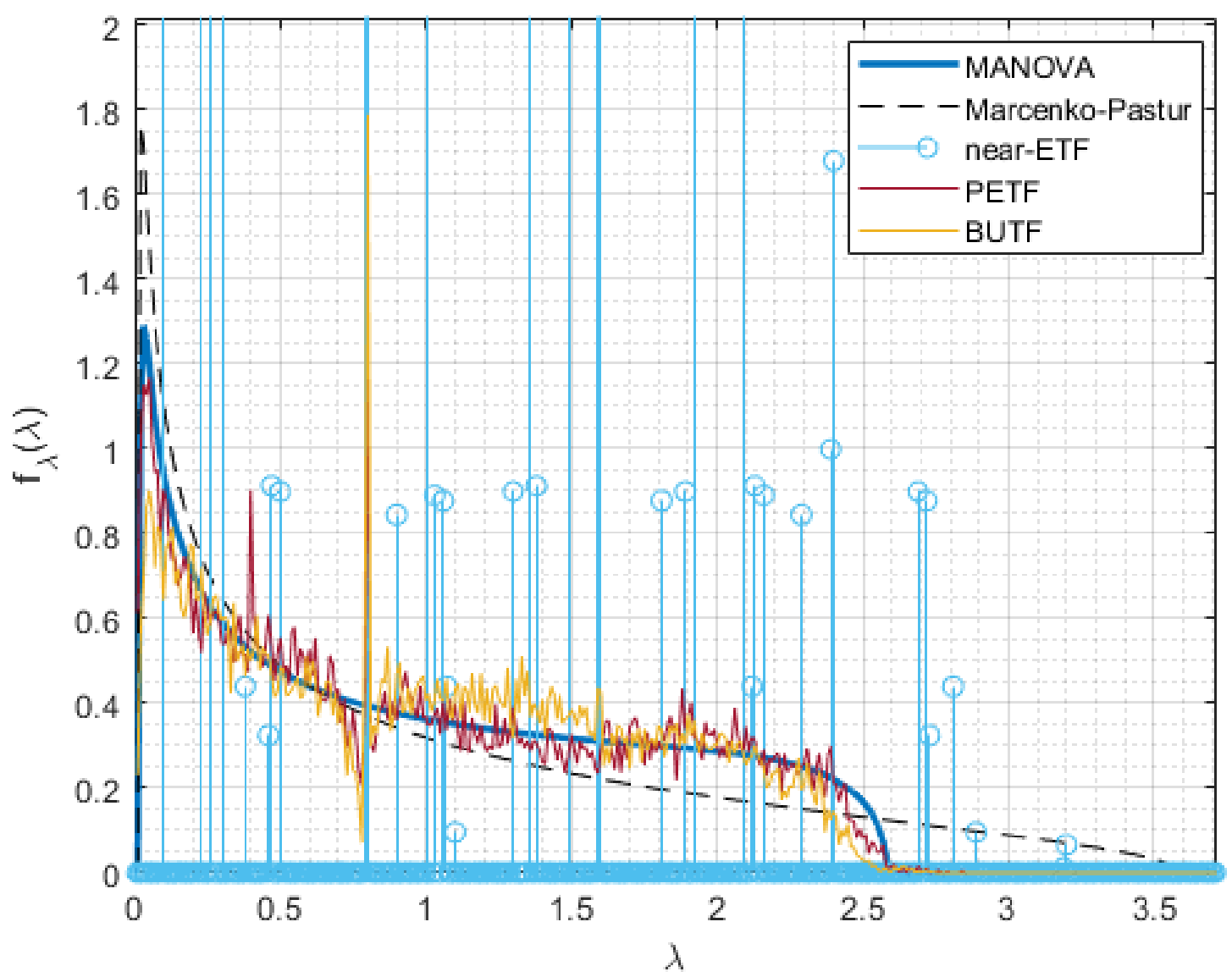

Figure 4: Eigenvalue distribution comparison for $N = 64, N_B = 16, \beta^{-1} = 1.25, \gamma = 0.31$

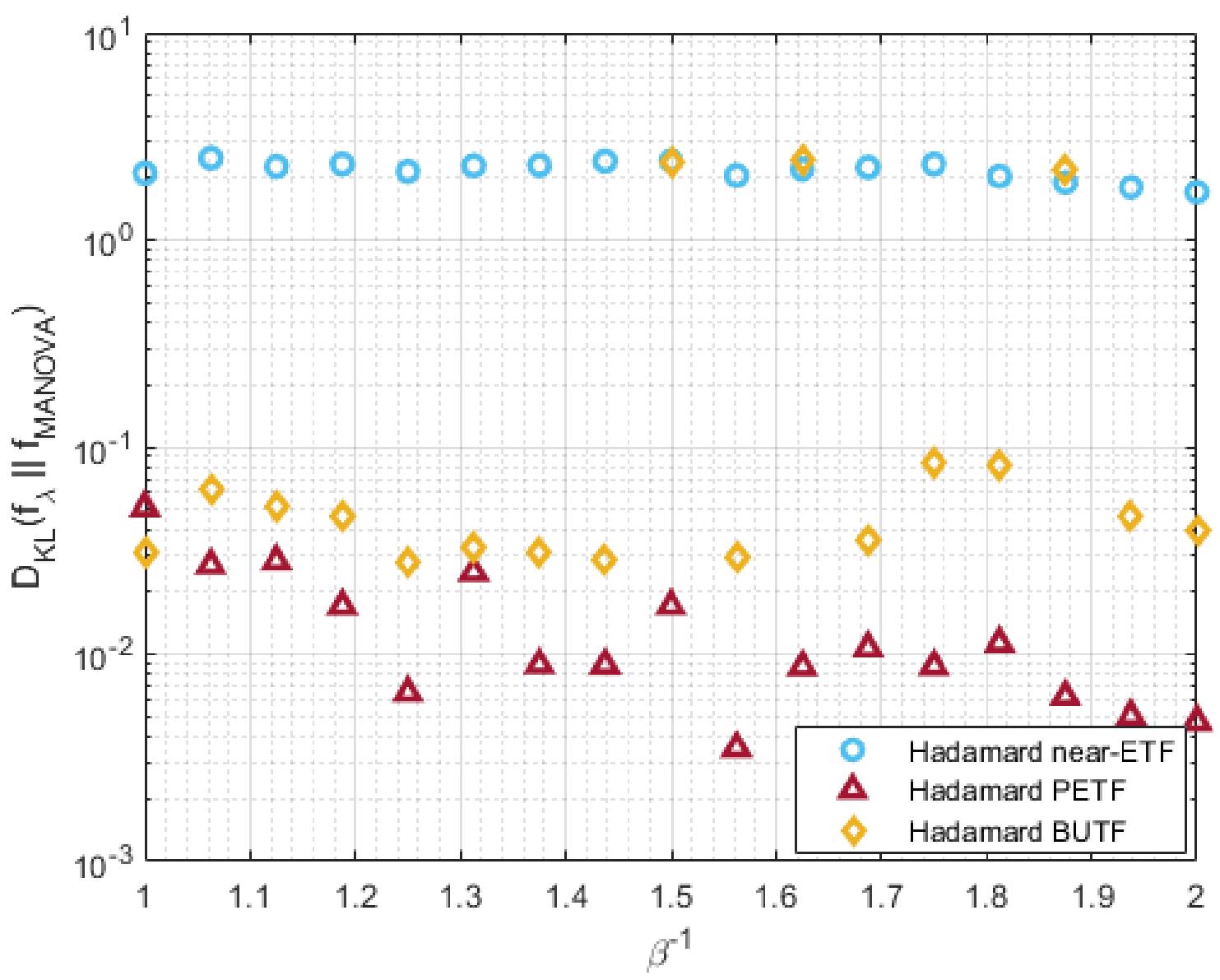

Figure 5: KL-Divergence comparison for $N = 64, N_B = 16, p = 0.25$

# V. PERFORMANCE OF BLOCK-ERASURE FRAMES

We present in this section performance of Hadamard frame codes in NOMA-CDMA and STC setups, including comparison of canonical near-ETF, PETF and BUTF. Figure 6 shows the average capacity in a NOMA-CDMA setup under the block-erasure channel for different values of $\beta^{-1}$ for BUTFs, PETFs and for canonical near-ETFs. It also contains reference performance of the MP [28] and MANOVA distributions (8), which are relevant for i.i.d frames and ETF, respectively, under uniform random erasures. Figure 7 shows the capacity outage probability for a rate equal to 98% of the average capacity for different values of $\beta^{-1}$. Additionally, in Figure 8 we show the average error probability upper bound (23) in STC setup for the different frames over different SNR values. Results show that canonical near-ETF frames are ill-suited for the block-erasure channel, and in some cases perform worse than i.i.d frames. In contrast, both PETF and BUTF frames have an average gain over canonical near-ETF of several bits/sec in a NOMA-CDMA setup or an average gain of up to 1 dB in the STC setup. They also show higher average capacity than the MANOVA benchmark performance. Furthermore, the capacity distribution of these frames has a lower variance, leading to a smaller capacity outage probability. The performance of BUTF and PETF is mostly similar, but we do see in some cases higher average capacity for BUTF compared to PETF, and lower variance of the capacity (or outage probability) for PETF compared to BUTF.

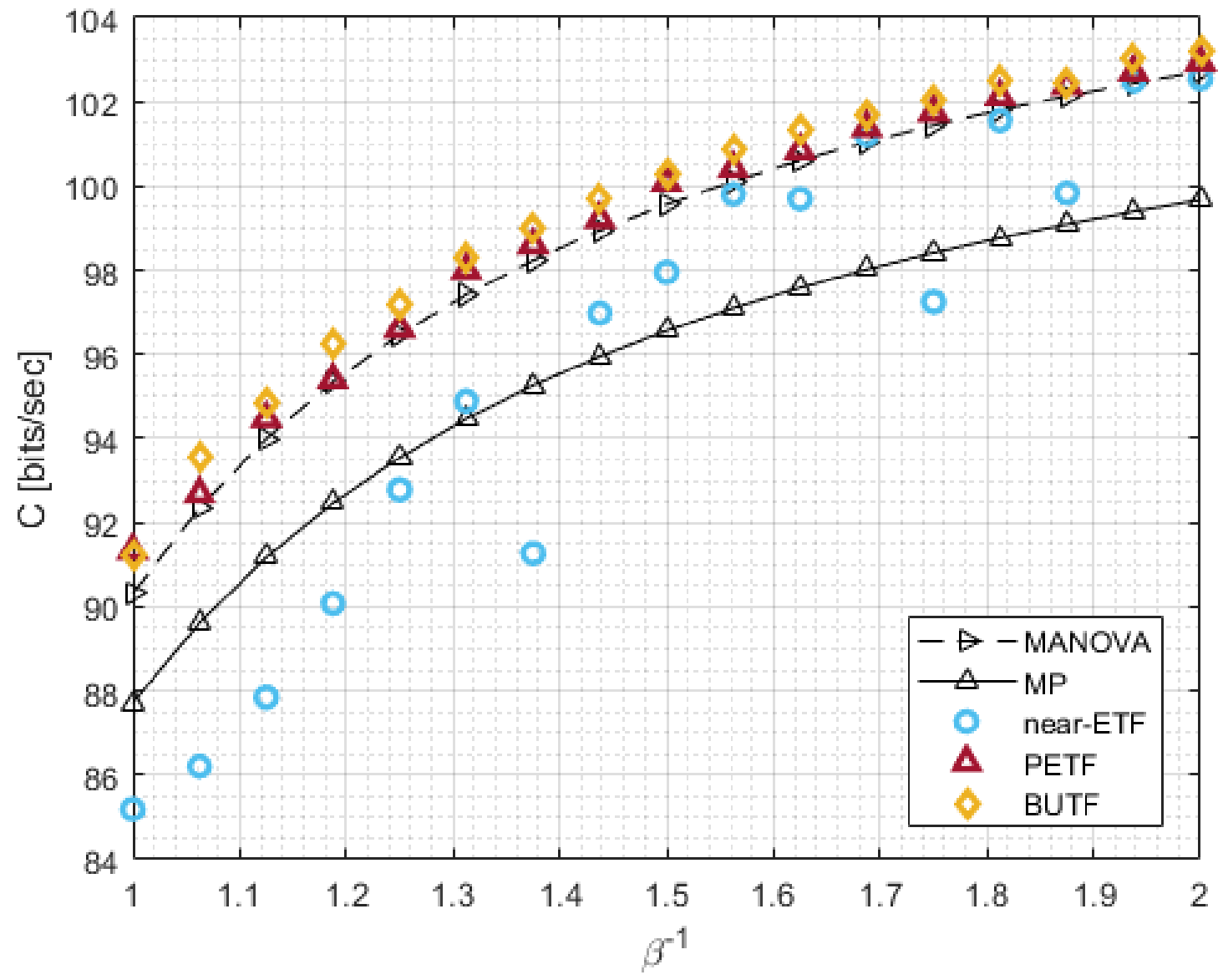

Figure 6: Average capacity [bits/sec] vs. $\beta^{-1}$ for $N = 64, N_B = 16, p = 0.25, SNR = 20\ dB$

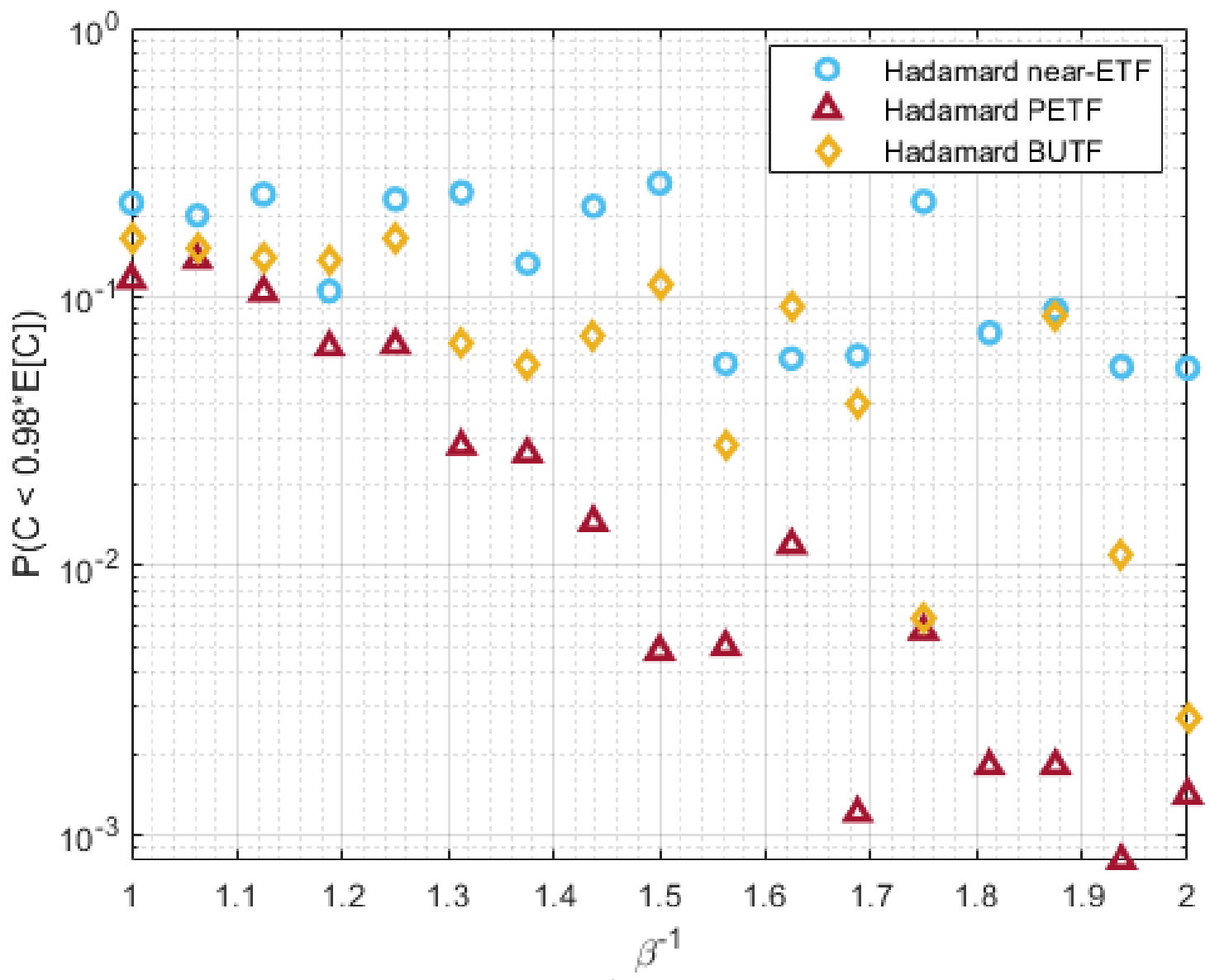

Figure 7: Capacity outage probability vs. $\beta^{-1}$ for $N = 64, N_B = 16, p = 0.25, SNR = 20\ dB$

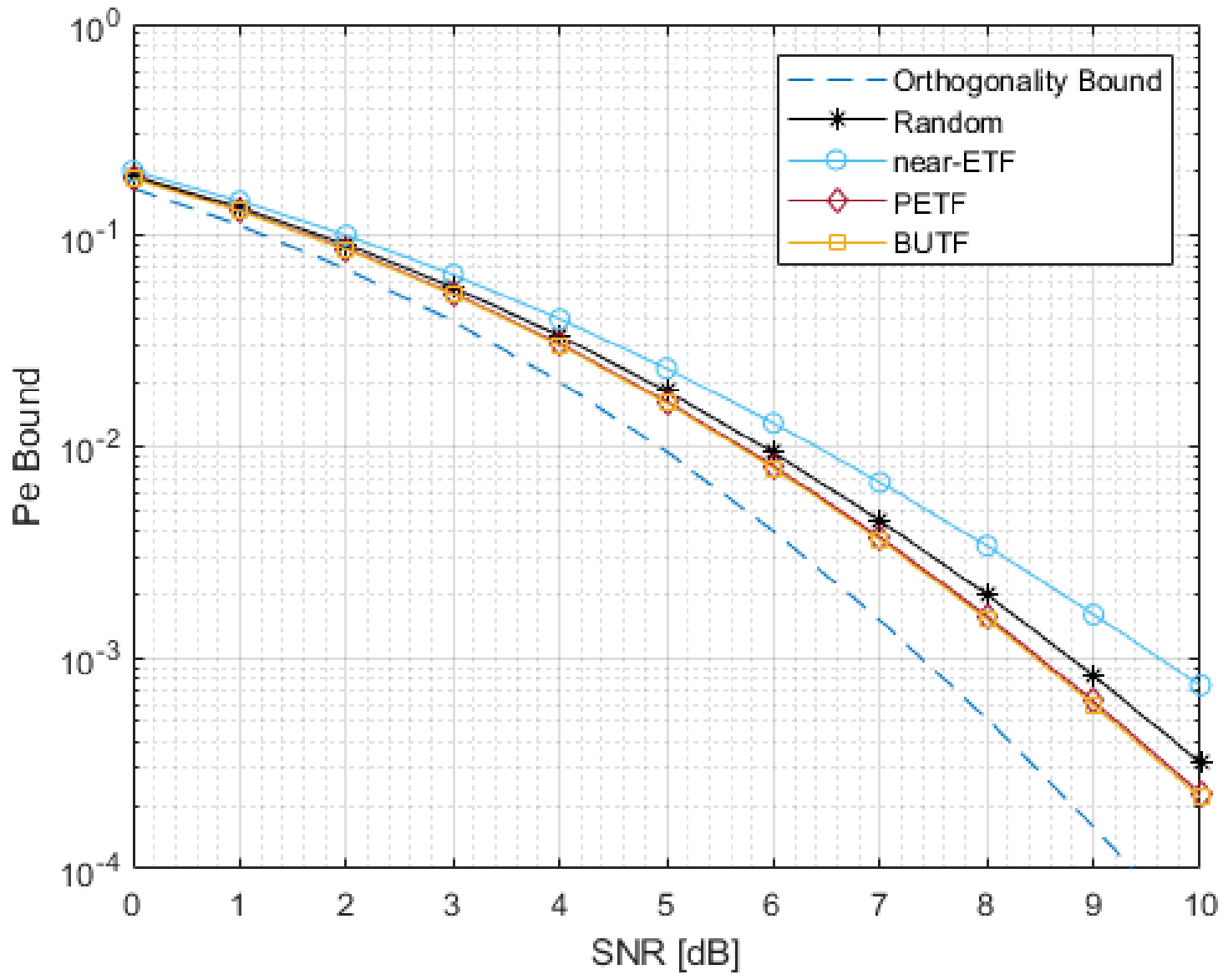

Figure 8: Average probability of error bound vs SNR [dB] for $N = 64, M = 8, N_B = 16, p = 0.12$

# VI. Conclusion and Discussion

We proposed to use frame codes over the block-erasure channel, a generalization of the uniform erasure channel, and demonstrated it in the NOMA-CDMA and STC setups. We showed empirically that under block erasures, a proper permutation of the vectors of ETF is essential to have good performance and exhibit a MANOVA spectrum. We also found by empirical means frames that outperform ETFs for block-erasures, and showed that a necessary condition for good block-erasure frame codes is having a low intra-block correlation. Both suggested frame codes show better results than canonical ETFs under a block-erasure channel, and provide higher average capacity, lower capacity outage probability and lower error probability.

We remain with some open questions: Is there are a fundamental difference in the asymptotic behavior in the case where the erasure block size $N_v$ increases linearly with $n$, versus the case where the erasure block size is fixed as $n \to \infty$? What is the asymptotic eigenvalue value distribution (of a subframe) in each of the two cases? Are the results sensitive to the alphabet of the frame (binary, discrete, real, complex)? What is the behavior of non-harmonic frames under block erasures? Another open problem is providing a rigorous proof for the necessary conditions on the correlation for good block erasure frames, as well as finding a sufficient condition. Furthermore, it is interesting whether such a proof will lead to the asymptotic optimality of the MANOVA spectrum in the block setup.

## תקציר

קודים אנלוגים מוסיפים יתירות על ידי הרחבת המימד תוך שימוש בפעולות ממשיות או קומפלקסיות. תורת המסגרות נותנת בסיס מתמטי לבניית קודים מסוג זה, עם שימושים נרחבים בתקשורת מרובת משתמשים בקידוד לא-אורתוגונלי (NOMA-CDMA), חישוב מבוזר, קידוד מקור באמצעות תיאורים מרובים (multiple description), קודי זמן-מרחב ועוד. מודל הערוץ שתואם לאפליקציות אלו הוא שילוב של רעש ומחיקות. לאחרונה, אנליזות הראו קישור מועיל בין התורה הספקטרלית של מטריצות אקראיות למסגרות הדוקות שוות-זווית (ETFs) תחת מחיקות אקראיות בהסתברות אחידה. בעבודה זו אנו מכלילים את המודל לעיל לערוץ בו המחיקות מקובצות בבלוקים. מודל זה מתאים במיוחד עבור NOMA-CDMA עם ריבוי אנטנות שידור לכל משתמש וקודי זמן מרחב כאשר יש קורלציה מרחבית גבוהה בין קבוצות שונות של אנטנות. אנו מציגים כאן שיטה לסדר מסגרות מסוג ETF כך שיתאימו למחיקה בבלוקים, ומציגים מסגרות בעלות קורלציה תוך-קבוצתית מינימלית בעלות ביצועיים טובים יותר מ-ETF בתנאים אלו.

אוניברסיטת תל-אביב

# קודי מסגרת לערוץ בלוק-מחיקה

איתמר יעקבי

מנחה:
פרופ' רם זמיר